\documentstyle[11pt,newpasp,twoside,epsf]{article}
\reversemarginpar

\begin{document}
\title{Slow pulsars from the STScI/NAIC drift scan search}
\author{K.~Xilouris}
\affil{Center Development Lab, NRAO, Charlottesville, VA 22903, USA}
\author{A.~Fruchter}
\affil{Space Telescope Science Institute, Baltimore, MD 21218, USA}
\author{D.R.~Lorimer, J.~Eder, A.~Vazquez}
\affil{Arecibo Observatory, HC3 Box 53995, Arecibo, PR 00612, USA}

\begin{abstract}
The unique sensitivity of the Arecibo telescope at 430 MHz motivated a
drift scan search of the visible sky divided up into eight declination
strips. Based on an analysis of the data collected in the region assigned
to us, eight new long-period pulsars have so far been discovered.
\end{abstract}

\section*{The STScI/NAIC drift scan search}

During the recent Gregorian upgrade of the Arecibo telescope,
considerable effort was put into drift scan searches of the Arecibo
sky ($-1^{\circ} < \delta < +39^{\circ}$) for new pulsars.  The
STScI/NAIC group was assigned declination strips centered at
1.5$^{\circ}$, 6.5$^{\circ}$, 11.5$^{\circ}$, 16.5$^{\circ}$,
21.5$^{\circ}$, 26.5$^{\circ}$, 31.5$^{\circ}$, and 36.5$^{\circ}$. A
list of 20 candidates was compiled from a search in these areas
between 1994 and 1998.  We have so far confirmed eight new pulsars as
a result of these observations.

The nominal parameters based on the confirmation observations are
summarized in Table 1.  Barycentric periods have uncertainties of
order one unit in the last digit quoted, while a conservative estimate
of the uncertainty in the dispersion measures (DM) is $\pm 10$ cm$^{-3}$
pc. The positions are presently uncertain by of order $\pm 5$ arcmin
in right ascension and declination --- equivalent to the half power 
beam size of the telescope at 430 MHz.

Although we presently have no long-term estimates of the flux
densities of the new pulsars, it is already clear that they are weak
sources with typical flux densities of order 0.5 to 1 mJy.  Some of
the initial detections were probably significantly facilitated by flux
amplifications due to interstellar scintillation.  Inferred 430-MHz
luminosities, based on their fluxes and dispersion measures estimates
range between 3 and 30 mJy kpc$^2$.  These pulsars, along with those
discovered by other groups during the Arecibo upgrade, should greatly
assist future statistical studies of the low end of the pulsar
luminosity function.  More accurate measurements of the flux
densities, as well as the spin and astrometric parameters for each
source are presently underway at Arecibo as part of a regular timing
program using the Penn State Pulsar Machine.

\begin{table}[hbt]
\begin{center}
\begin{tabular}{llllll}
\hline
PSR     &  R.A.  &  Decl.              & Period & Epoch & DM \\
        &\multicolumn{2}{c}{(J2000)}   &  (sec) & (MJD) & cm$^{-3}$ pc\\
\hline
J0137+16&  01:37:31&   +16:55 &  0.41477  &  51264&   26      \\
J0329+16&  03:29:13&   +16:54 &  0.8933   &  51257&   35      \\
\\
J1549+21&  15:49:41&   +21:14 &  1.262    &  51335&   55      \\
J1822+11&  18:22:17&   +11:22 &  1.787    &  51261&  112      \\
\\
J1838+16&  18:38:52&   +16:53 &  1.902    &  51261&   36      \\
J1849+06&  18:49:07&   +06:07 &  2.219    &  51258&  236      \\
\\
J1905+06&  19:05:21&   +06:23 &  0.9897   &  51261&  262      \\
J2040+16&  20:40:13&   +16:54 &  0.8656   &  51261&   51    \\
\hline
\end{tabular}
\end{center}
\caption[]{
Parameters of the eight newly-discovered pulsars based on the
discovery and confirmation observations.
}
\end{table}

\begin{figure}[hbt]
\setlength{\unitlength}{1in}
\begin{picture}(0,3.5)
\put(0,4){\includegraphics{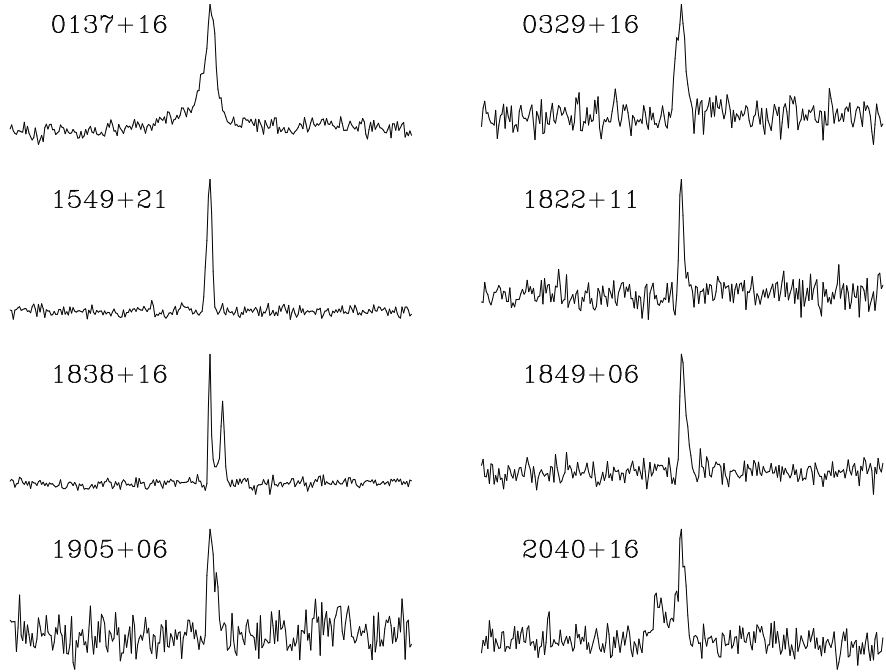}}
\end{picture}
\caption[]{
Integrated 430-MHz pulse profiles for the eight newly-discovered
pulsars. Each profile represents $360^{\circ}$ of rotational phase.
}
\end{figure}

\end{document}